\documentclass[12pt, a4paper]{article}
\usepackage{a4wide,amssymb,amsmath,amscd}

\newcommand{\T}{{\mathbb T}}

\newcommand{\R}{{\mathbb R}}

\begin{document}

\topmargin -2pt

\headheight 0pt

\topskip 0mm \addtolength{\baselineskip}{0.20\baselineskip}
\begin{flushright}
{\tt KIAS-P07077}
\end{flushright}

\vspace{5mm}

\begin{center}
{\Large \bf Noncommutative Superspace  and \\ Super Heisenberg Group} \\

\vspace{10mm}

{\sc Ee Chang-Young}\footnote{cylee@sejong.ac.kr}\\
{\it Department of Physics, Sejong University, Seoul 143-747, Korea}\\


\vspace{5mm}

{\sc Hoil Kim}\footnote{hikim@knu.ac.kr}\\

{\it Department of Mathematics, Kyungpook National University,\\
Taegu 702-701, Korea}\\

\vspace{2mm}

and \\

\vspace{2mm}

{\sc Hiroaki Nakajima}\footnote{nakajima@skku.edu}\\

{\it Department of Physics and Institute of Basic Science, \\
Sungkyunkwan University, Suwon 440-746, Korea}\\

\vspace{10mm}

{\bf ABSTRACT} \\
\end{center}

\vspace{2mm}

\noindent
In this paper, we consider noncommutative superspace in relation with
super Heisenberg group.
We construct a matrix representation of super Heisenberg group and
apply this to the two-dimensional deformed $\mathcal{N}=(2,2)$ superspace
that appeared in string theory.
We also construct a toy model for non-centrally extended
`super Heisenberg group'.
\\



\thispagestyle{empty}

\newpage
\section{Introduction}

\noindent

Noncommutative geometry\cite{conn} naturally appears in string
theory \cite{jp,gsw}. Low-energy effective theory of D-branes in a
background NSNS $B$-field becomes the noncommutative field theory
where the spacetime coordinates $x^{\mu}$ are noncommutative
$[x^{\mu}, x^{\nu}] \neq 0$ \cite{cds,ChuHo,SeWi}. On the other
hand, when we turn on the background RR field, low-energy
effective theory of D-branes becomes the field theory on
non(anti)commutative superspace of which the fermionic coordinate
$\theta^{\alpha}$ has nontrivial commutation relation
$\{\theta^{\alpha}, \theta^{\beta}\} \neq 0$
\cite{Klemm:2001yu,OoVa,de Boer:2003dn,Se,BeSe,ty,flm}. Gauge theories on
non(anti)commutative superspace are studied extensively
\cite{ArItOh,FeIvLeSoZu,Ito:2005jj,CaIvLeQu}. In the case of
constant $B$-field, the algebra of the coordinates becomes
$[x^{\mu},x^{\nu}] = \textrm{const.}$, which is called Heisenberg
algebra (or Weyl algebra as an universal enveloping algebra of
Heisenberg algebra). Heisenberg group
\cite{mumford,sthan98,jrosenberg} is constructed from the
Heisenberg algebra by exponential mapping. If we include both of
background NSNS and RR fields, it is expected that super
Heisenberg group \cite{lo94} would appear.

Heisenberg group and its Schr\"odinger representation are defined
rigourously in mathematics from the motivation of quantum
mechanics \cite{jrosenberg}. Heisenberg group is constructed as a
central extension of a symplectic vector space and its matrix
representation is constructed by triangular matrices. The
commutation relation $[x^{\mu},x^{\nu}] = \textrm{const.}$ is
regarded as the operator representation (Schr\"odinger
representation) of Heisenberg group. The matrix representation of
Heisenberg group is useful to construct noncommutative tori
\cite{cr1987,rief88,rs98} and quantum theta-functions
\cite{manin1,manin1-tr,manin2,manin3,ek06,ek07}.
Here noncommutative tori are defined by the commutator relation
coming from the cocycle condition used in Heisenberg group.
In this paper, for the anologue of bosonic Heisenberg group,
we find the representation of
super Heisenberg group from the supermatrices which
are also triangular up to exchange of their rows and columns.
This supermatrix representation is also applicable to
the superspace deformed by
noncommutive and non-anticommutative parameters,
and certain cases of non-central extensions. As in the bosonic
case, the supermatrix representation of super Heisenberg group is
useful to construct noncommutative supertori and
quantum super theta-functions \cite{EeKiNa}. Understanding in this
direction will be necessary for investigating the properties of supersymmetric
field theories such as soliton solutions on noncommutative
supertorus.

This paper is organized as follows. In section 2, we
review the basics of noncommutative space and the construction of
the Heisenberg group.  We also explain the relations between
the Heisenberg algebra, the operator representation of the Heisenberg
group, and the corresponding noncommutative space.
In section 3, we construct
the super Heisenberg group and its supermatrix and operator representations
extending the relations known in the bosonic case. In
section 4, we consider two types of deformed superspaces in
relation with super Heisenberg group; a two-dimensional superspace
deformed by noncommutative and non-anticommutative parameters, and
a toy model of non-centrally extended `super Heisenberg group'. We
conclude in section 5.
\\


%
\section{Noncommutative space and Heisenberg group}



\subsection{Noncommutative space}\label{heisgp}
Noncommutative space\footnote{
In general, one can consider many kinds of noncommutative spaces.
But we use the word \textit{noncommutative space} in the sence of \eqref{NC}.
}
is defined as a space on which coordinates
$X^{\mu}\ (\mu=1,2,\ldots,2n)$\footnote{
If there are odd number of the coordinates, one of them can commute with
all other coordinates.
}
satisfies the commutation relation
\begin{equation}
\bigl[X^{\mu}, X^{\nu}\bigr]=i\varTheta^{\mu\nu}.\label{NC}
\end{equation}
Here $\varTheta^{\mu\nu}$ is a constant.
Without loss of generality, $\varTheta^{\mu\nu}$ can take
the block-diagonal form
\begin{equation}
\varTheta^{\mu\nu}=
\left(\begin{array}{cc|cc|cc}
 & \varTheta^{12} & \multicolumn{4}{c}{} \\
-\varTheta^{12} & & \multicolumn{4}{c}{} \\
\cline{1-4}
 & & & \varTheta^{34} & & \\
 & &-\varTheta^{34} & & & \\
\cline{3-6}
 \multicolumn{3}{c}{} & & \ddots & \\
 \multicolumn{3}{c}{} & & & \ddots \\
\end{array}\right).
\end{equation}
Then the commutation relation becomes
\begin{equation}
\bigl[X^{2i-1}, X^{2i}\bigr]=i\varTheta^{2i-1,2i},\quad i=1,2,\ldots,n.
\end{equation}
In this basis, we can use the representation of $X^{\mu}$ as
\begin{equation}
X^{2i-1}=\sqrt{\varTheta^{2i-1,2i}}\,s^{i}\,\cdot\,,\quad
X^{2i}=-i\sqrt{\varTheta^{2i-1,2i}}\frac{\partial}{\partial s^{i}}.
\end{equation}

The algebra of the functions on
noncommutative space is equivalent to the algebra of the functions on
commutative space 
with the noncommutative product, which is called Moyal product:
\begin{equation}
F(X)\ast G(X)=\exp\biggl(\frac{i}{2}\varTheta^{\mu\nu}
\frac{\partial}{\partial X^{\mu}}\frac{\partial}{\partial X'{}^{\nu}}
\biggr)F(X)G(X^{\prime})
\biggr|_{X'{}^{\mu}=X^{\mu}}.
\end{equation}

\subsection{Heisenberg group}\label{heisgp}

\noindent
 We define a Heisenberg group, $\mathit{Heis}({\R}^{2n},\psi)$.
As a set $\mathit{Heis}({\R}^{2n},\psi)$ is $U(1) \times \R^{2n}$.
For $t, t' \in U(1)$, and $(x,y),(x',y') \in \R^{2n}$, we define
$(t,x,y),(t',x',y') \in$ $\mathit{Heis}({\R}^{2n},\psi)$,
\begin{equation}
(t,x,y) \cdot (t',x',y') = (t+t' + \psi(x,y;x',y'), x+x',y+y'),
\label{multi}
\end{equation}
where $  \psi:~~ {\R}^{2n} \times {\R}^{2n} \longrightarrow \R,$ satisfies
the cocycle condition
\begin{equation}
\psi(x,y;x',y')\psi(x+x',y+y';x'',y'')=
\psi(x,y;x'+x'',y'+y'')\psi(x',y';x'',y''),
\end{equation}
which is a necessary and sufficient condition for the multiplication
to be associative.
There is an exact sequence
\begin{equation}
0 \rightarrow \mathbb{R} \mathop{\rightarrow}^{i}
\mathit{Heis}(\mathbb{R}^{2n},\psi) \mathop{\rightarrow}^{j}
\mathbb{R}^{2n} \rightarrow 0
\end{equation}
called a central extension, with the inclusion $i(t)=(t,0,0)$ and the projection $j(t,x,y)=(x,y)$,
where $i(\mathbb{R})$ is the center in
$\mathit{Heis}({\R}^{2n},\psi)$.

 We now introduce two representations of this
Heisenberg group $\mathit{Heis}({\R}^{2n},\psi)$.
One is a matrix representation and
the other is an operator representation.

 First we introduce the matrix representation
$\mathit{Heis}({\R}^{2n},\psi)\longrightarrow \mathit{Mat}_{(n+2,n+2)}({\R})$ in two ways.
One of them is given by
\begin{equation}
\label{mat1}
(t,x,y) \longrightarrow
 \begin{pmatrix}
 1 & x & t \\
 0 & 1 & y \\
 0 & 0 & 1
 \end{pmatrix} = \mathcal{M}(t,x,y).
\end{equation}
In this case $\psi(x,y;x',y')=xy'$.

For the other one, we  use the Lie algebra and Lie group approach. Let
\begin{equation}
M(t,x,y) =
 \begin{pmatrix}
 0 & x & t \\
 0 & 0 & y \\
 0 & 0 & 0
 \end{pmatrix}.
\end{equation}
Then
\begin{equation}
\mathcal{M}(t,x,y) =
 \begin{pmatrix}
 1 & x & t \\
 0 & 1 & y \\
 0 & 0 & 1
 \end{pmatrix} =
 \begin{pmatrix}
 1 & 0 & 0 \\
 0 & 1 & 0 \\
 0 & 0 & 1
 \end{pmatrix} +
 \begin{pmatrix}
 0 & x & t \\
 0 & 0 & y \\
 0 & 0 & 0
 \end{pmatrix} = I + M(t,x,y).
\end{equation}
For the second matrix representation, we define
\begin{equation}
\label{mat2}
\pi (t,x,y) := e^{M(t,x,y)} = I + M + M^2/2 + \cdots =
 \begin{pmatrix}
 1 & x & t + \frac{xy}{2} \\
 0 & 1 & y \\
 0 & 0 & 1
 \end{pmatrix}.
\end{equation}
Then,
$\mathit{Heis}({\R}^{2n},\psi)$ with the cocycle $\psi(x,y;x',y')=\frac{1}{2}(xy'-yx')$
is isomorphically embedded in $\mathit{Mat}_{(n+2,n+2)}(\R)$.
In this case, $\psi$ becomes a symplectic form,
and then this representation can be extended to
general symplectic vector spaces%
\footnote{In general, $(x,y)$ is a Darboux pair of
the symplectic vector space and $\psi$ is the symplectic form.}.
We can easily show that both $\mathcal{M}$ in (\ref{mat1}) and $\pi$ in (\ref{mat2}) are group
homomorphisms, which shows that they are matrix representations of the Heisenberg group.

The Lie algebra introduced in the second matrix representation is a vector space isomorphic to $\R \times \R^{2n}$
generated by
$q_{i}, ~ p_{i} ~ (i=1, \dots, n),$ and $r$ such that
\begin{equation}
\label{cmlie}
[q_{i}, p_{j}]=\delta_{ij}r, \quad \text{others}=0.
\end{equation}
The above Lie algebra is the so-called Heisenberg algebra, $h(2n)$.
A general element of this Heisenberg algebra $h(2n)$ is of the form $x\cdot q+y\cdot p+tr$.
In the second matrix representation,
the generators are mapped as follows
\begin{align}
q_{i}\to Q_{i}&=
\begin{pmatrix}
0 & \ldots & 1 & \ldots & 0 \\
  & \ddots &   &        & \vdots \\
  &        & \ddots &   & \vdots \\
  &        &        & \ddots & \vdots \\
  &        &        &        & 0 \\
\end{pmatrix},\quad
p_{i}\to P_{i}=
\begin{pmatrix}
0 & \ldots & \ldots & \ldots & 0 \\
  & \ddots &   &        & \vdots \\
  &        & \ddots &   & 1 \\
  &        &        & \ddots & \vdots \\
  &        &        &        & 0 \\
\end{pmatrix},\notag\\
r\to R&=
\begin{pmatrix}
0 & \ldots & \ldots & \ldots & 1 \\
  & \ddots &   &        & \vdots \\
  &        & \ddots &   & \vdots \\
  &        &        & \ddots & \vdots \\
  &        &        &        & 0 \\
\end{pmatrix},
\end{align}
where the component 1 in $Q_{i}$ appears only in the $(i+1)$-th column of the first row,
and the component 1 in $P_{i}$ appears only in the $(i+1)$-th row of the last column.
In this notation, $\pi(t,x,y)$ in \eqref{mat2} can be expressed as
\begin{equation}
\pi(t,x,y)=e^{x\cdot Q + y\cdot P + t R}.
\label{mat3}
\end{equation}
Note that $\pi (t,x,y) \cdot \pi (t',x',y')$ can be
expressed as
\begin{equation}
e^{M(t,x,y)} \cdot e^{M(t',x',y')} = e^{M(t,x,y)+M(t',x',y')+
\frac{1}{2}[M(t,x,y),M(t',x',y')] + \cdots } ~~~
\end{equation}
and we will use this property in our computations.

Next, we explain the Schr$\ddot{\rm o}$dinger representation which is an operator representation
on $L^2(\R^n)$ of the Heisenberg group, where $L^2 (\R^n)$ is the completion of the Schwarz space on $\R^n$.
 For this we reexpress the commutation relation of the Heisenberg algebra (\ref{cmlie})
 with new generators $X_i, ~ Y_i,$ and $I$
 by the following map
 \begin{equation}
 \label{opmap}
 q_i \longrightarrow i X_i , ~ p_i  \longrightarrow i Y_i, ~ r \longrightarrow -i  I .
 \end{equation}
 Then we have
\begin{equation}
\label{cmheis}
[X_i, Y_j] = i \delta_{ij} I , ~~ {\rm others}=0.
\end{equation}
A representation for $X_i, Y_i, I$ as operators can be given by
\begin{equation}
(X_i f)(s)=s_i f(s),\quad
(Y_i f)(s)=-i\frac{\partial f}{\partial s_i}(s), \quad
I = {\rm identity},
\end{equation}
where $s=(s_1,\cdots,s_n)$.
Note that the above  $X_i, Y_i$ can be regarded as the noncommutative coordinates
$X^\mu$ introduced in the previous subsection.

 For the operator representation of the first matrix representation $\mathcal{M}(t,x,y)$ with
the cocycle $\psi(x,y;x',y')=xy'$,
 we send $(t,x,y) \longrightarrow \chi(t)e(x)d(y)$, where
\begin{align}
\chi(t):~f(s) & \longrightarrow e^{- i t } f(s), \nonumber \\
e(x):~f(s) & \longrightarrow e^{ i x \cdot X } f(s)
=e^{ i x \cdot s } f(s),\\
d(x):~f(s) & \longrightarrow e^{i y \cdot Y } f(s) = f(s+y). \nonumber
\end{align}

For the operator representation of the second matrix representation
 $\pi(t,x,y)$ with the cocycle
 $\psi(x,y;x',y')=\frac{1}{2}(xy'-yx')$,
we send $(t,x,y)$ to $U(t,x,y)=\chi(t+\frac{xy}{2})e(x)d(y)$.
Note that $U(t,x,y)$
can be rewritten as
\begin{equation}
U(t,x,y)=\exp\bigl[i(x\cdot X+y\cdot Y-t I)\bigr],
\end{equation}
and this corresponds to $\pi(t,x,y)$ in (\ref{mat3}) via (\ref{opmap})

The noncommutative parameter is
\begin{equation}
\label{copsi}
 \varTheta(x,y;x',y')=\psi(x,y;x',y')-\psi(x',y';x,y)=xy'-yx'
\end{equation}
for both cases.
In the second case, $\psi(x,y;x',y')=\frac{1}{2}(xy'-yx')$, so that
$\varTheta=2\psi$.

In the later part of our work, we will use the second operator representation,
which can be mapped into a noncommutative space
as we mentioned above,
is irreducible and unitary.
And due to Stone-von Neumann-Mackey theorem, this is unique  up to isomorphism.

Now we explain Stone-von Neumann-Mackey theorem
\cite{mumford,sthan98,jrosenberg}. First we define
$e:\mathbb{R}^{2n} \times \mathbb{R}^{2n} \rightarrow \mathbb{R}$
by
$e(z_1,z_2)=\tilde{z_1}\tilde{z_2}\tilde{z_1}^{-1}\tilde{z_2}^{-1}$
where $z_1,z_2 \in \mathbb{R}^{2n}$, and $\tilde{z_i}$ is a
lifting of $z_i$ which means $j(\tilde{z_i})=z_i$. Then $e$ is
well defined independent of the choice of $\tilde{z_i}$ and is a
skew-symmetric pairing.
In our case, $e$ corresponds to $\varTheta$ in (\ref{copsi}).
If for some subgroup $H \subset
\mathbb{R}^{2n}$, $e|_{H \times H}=0$, then $H$ is called an
isotropic subspace(or Largrangian). If $H$ is maximal among those
isotropic subspaces, it is called a maximal isotropic subspace.
For example, $H$ can be the $x$-space or the $y$-space or some
combination.
Now we state the theorem \cite{mumford}.\\
\textbf{Theorem}(Stone, von Neumann, Mackey)
\\Let $G=\mathit{Heis}(\mathbb{R}^{2n},\psi)$ be a Heisenberg group. Then
\\1) $G$ has a unique irreducible unitary representation
\begin{equation}
U : G \longrightarrow Aut(\mathcal{H}_0)
\end{equation}
such that $U_t=e^{-it}\cdot I$ for all $t\in \mathbb{R}$.
\\2) For all maximal isotropic subgroup $H \subset \mathbb{R}^{2n}$,
and a lifting $\sigma(z)=(\alpha(z),z)$ of $j$ over $H$, where $\alpha$
is a group homomorphism from $H$ to $\mathbb{R}$,
this representation may be realized by
\\$\mathcal{H}_0= \{ \mbox{ measurable function }f:\mathbb{R}^{2n}
\rightarrow \mathbb{C}$, \mbox{ such that }
\begin{align}
\text{a)} &\quad f(z+h)=\alpha(h)^{-1} \psi(h,z)^{-1} f(z), \quad
\forall h \in H,
\nonumber\\
\text{b)} &\quad \int |f(z)|^2 dz < \infty. \quad
U_{(t,z')}f(z)=e^{-it}\psi(z,z')f(z+z'),\quad
\forall z' \in \mathbb{R}^{2n} \}.
\nonumber
\end{align}
3) All representations $(U,\mathcal{H})$ such that $U_t=e^{-it}\cdot I$,
all $t \in \mathbb{R}$, are isomorphic to
$\mathcal{H}_0 \otimes \mathcal{H}_1$, and
$G$ acting trivially on $\mathcal{H}_1$.
\\


\section{Super Heisenberg group}\label{spheis}

Now, we consider the extension of the
work in the previous section to
the super case.
As in the bosonic case, we define the super Heisenberg group, $\mathit{sHeis}({\R}^{2n|2m},\psi)$,
as follows.
 For $t, t' \in U(1)$, and $(x, \alpha), (y, \beta), (x',\alpha'), (y', \beta') \in {\R}^{n|m} ,$ we define
 $(t,x,y, \alpha, \beta),(t',x',y', \alpha', \beta') \in$ $\mathit{sHeis}({\R}^{2n|2m},\psi)$
 such that
\begin{equation}
(t,x,y, \alpha, \beta) \cdot (t',x',y', \alpha', \beta')  =
(t+t' + \psi(x,y, \alpha, \beta;x',y',\alpha', \beta' ), x+x',y+y', \alpha + \alpha' , \beta + \beta'),
\label{multi}
\end{equation}
where $  \psi:~~ {\R}^{2n|2m} \times {\R}^{2n|2m} \longrightarrow \R,$ satisfies
the cocycle condition
\begin{eqnarray}
&& \psi(x,y, \alpha, \beta;x',y',\alpha', \beta')\psi(x+x',y+y',\alpha + \alpha' , \beta + \beta' ;x'',y'',\alpha'', \beta'' ) \nonumber \\
&&=
\psi(x,y, \alpha, \beta;x'+x'',y'+y'',\alpha' + \alpha'' , \beta' + \beta'')\psi(x',y', \alpha', \beta';x'',y'',\alpha'', \beta'' ),
\end{eqnarray}
 a necessary and sufficient condition for associative multiplication.
Now, there is an exact sequence
\begin{equation}
0 \rightarrow \mathbb{R} \mathop{\rightarrow}^{i}
\mathit{sHeis}(\mathbb{R}^{2n|2m},\psi) \mathop{\rightarrow}^{j}
\mathbb{R}^{2n|2m} \rightarrow 0,
\end{equation}
a central extension, with the inclusion $i(t)=(t,0)$, the projection $j(t,z)=z, ~ z \in \mathbb{R}^{2n|2m}$ ,
where $i(\mathbb{R})$ is the center in
$\mathit{sHeis}({\R}^{2n|2m},\psi):= sH(2n|2m)$.
As in the bosonic case, we can introduce two types of matrix representations and the corresponding operator
representations for the super Heisenberg group.

 First, we consider the matrix representations,
 $sH(2n|2m)
\longrightarrow Mat_{(n+2|m,~n+2|m)}({\R})$.
The first matrix representation is given by
\begin{align}
(t,x,y,\alpha,\beta) \longrightarrow
\left(
\begin{array}{ccc|c}
1 & x & t &\alpha \\
0 & 1 & y & 0 \\
0 & 0 & 1 & 0 \\
\hline
0 & 0 & \beta & 1
\end{array}
\right)
  = \mathcal{M}(t,x,y,\alpha,\beta) .
  \label{heism}
\end{align}
In this case $\psi(x,y,\alpha,\beta;x',y',\alpha',\beta')=xy'+\alpha\beta'$.

For the Lie algebra and Lie group approach, we let
\begin{equation}
 \mathcal{M}(t,x,y,\alpha,\beta)  := I + M(t,x,y,\alpha,\beta),
  \end{equation}
where
\begin{equation}
 M(t,x,y,\alpha,\beta) = \left(
\begin{array}{ccc|c}
0 & x & t & \alpha \\
0 & 0 & y & 0 \\
0 & 0 & 0 & 0 \\
\hline
0 & 0 & \beta & 0
\end{array}
\right).
 \end{equation}
We then define
\begin{equation}
 \pi (t,x,y,\alpha,\beta) := e^{M(t,x,y,\alpha,\beta)}
 = I + M + M^2/2 + \cdots
 =
 \left(
\begin{array}{ccc|c}
1 & x & t + \frac{xy +\alpha\beta}{2} &\alpha \\
0 & 1 & y & 0 \\
0 & 0 & 1 & 0 \\
\hline
0 & 0 & \beta & 1
\end{array}
\right).
\end{equation}
In this case  $\psi(x,y,\alpha,\beta;x',y',\alpha',\beta')=
\frac{1}{2}(xy'-yx' + \alpha\beta' +\beta\alpha')$.

The super Lie algebra introduced in the second matrix representation is a vector space isomorphic to $\R \times \R^{2n|2m}$
generated by
$q_{i}$, $p_{i}, ~\xi_a, ~ \lambda_a $ $(i=1,\ldots, n, ~  a= 1,\ldots, m),$ and $r$ such that
\begin{equation}
\label{cmsplie}
[q_{i}, p_{j}]=\delta_{ij}r, \quad \{\xi_a , \lambda_b \}=\delta_{ab}r,  \quad   \text{others}=0.
\end{equation}
The above super Lie algebra is the so-called super Heisenberg algebra, $sh(2n|2m)$.
A general element of this super Heisenberg algebra $sh(2n|2m)$ is of
the form $x\cdot q+y\cdot p+ \alpha \cdot \xi   +  \beta \cdot \lambda  + tr$.
In this representation,
the generators are mapped as follows similar to the bosonic case:
\begin{eqnarray}
&& q_{i} \rightarrow
\left(
\begin{array}{ccc|c}
0 &  e_i^t & 0 & 0 \\
0 & 0 &   0 &0 \\
 0& 0  & 0 & 0 \\
\hline
0 & 0 & 0 & 0
\end{array}
\right),
\quad
p_{i} \rightarrow
\left(
\begin{array}{ccc|c}
0 &  0 & 0 & 0 \\
0 & 0 &  e_i &0 \\
 0& 0  & 0 & 0 \\
\hline
0 & 0 & 0 & 0
\end{array}
\right),
\quad
r \rightarrow
\left(
\begin{array}{ccc|c}
0  & 0& 1 & 0 \\
0 &0 &  0 & 0 \\
0 & 0 & 0 & 0 \\
\hline
0 & 0 & 0 & 0
\end{array}
\right), \nonumber
\\
&& \xi_{a} \rightarrow
\left(
\begin{array}{ccc|c}
0 &  0 & 0 & e_a^t \\
0 & 0 &  0 &0 \\
 0& 0  & 0 & 0 \\
\hline
0 & 0 & 0 & 0
\end{array}
\right),
\quad
\lambda_{a} \rightarrow
\left(
\begin{array}{ccc|c}
0 &  0 & 0 & 0 \\
0 & 0 &  0 &0 \\
 0& 0  & 0 & 0 \\
\hline
0 & 0 & e_a & 0
\end{array}
\right),
\end{eqnarray}
where $e_i$ is the column vector in which the $i$-th component is 1 and the others vanish, and the same for $e_a$.

Let $V$ be a real super vector space of dimension $2n|2m$ with non-degenerate skew-symmetric form $(~,~)$.
Let $q_i,~ p_i,~\xi_a,~\lambda_a ,$ for $ i=1,\ldots, n ~ $ and $ a= 1,\ldots, m ,$ be a basis of $V$ such that the matrix of $(~,~)$
with respect to this basis is
\begin{equation}
\phi =
 \begin{pmatrix} 0 & 1 &  & \textbf{0}  \\
                 -1 & 0 & & &    \\
                    &  & 0 & 1    \\
                 \textbf{0} & & 1 & 0
             \end{pmatrix}
\end{equation}
i.e.
$(q_i,q_j)=(p_i,p_j)=(\xi_a,\xi_b)=(\lambda_a,\lambda_b)=0$
and $(q_i,p_j)=\delta_{ij}, ~(\xi_a,\lambda_b)= \delta_{ab}$.
Here $\phi$
corresponds to the map $e$ which we introduced for
Stone-von Neumann-Mackey theorem
 in the bosonic case.
 So $V=V_{\bar{0}}\otimes
V_{\bar{1}}$, where $\{q_i,p_i\}$ is a basis for $V_{\bar{0}}$ and
$\{\xi_a, \lambda_a\}$ a basis for $V_{\bar{1}}$.
The super Heisenberg algebra $sh(V)$ can be constructed as a central
extension of the Abelian Lie superalgebra $V$ by an even generator
$r$ \cite{lo94}. We have an exact sequence  $0 \to \mathbb{R}\cdot r \to sh(V)
\to V \to 0$, and the Lie bracket is defined by $[u,v]=(u,v)r,
\forall u,v \in V$. Then,  $sh(V)=sh(2n|2m)$ .

The operator representation  can be given as in the bosonic case.
The target space for the super Schr${\rm \ddot{o}}$dinger
representation is
$L^2(\mathbb{R}^{n|m}):=L^2(\mathbb{R}^n) \otimes
\bigwedge^*(\mathbb{R}^m)^*$, which is the completion of the Schwarz space
$S(\mathbb{R}^{n|m}):=S(\mathbb{R}^n)\otimes
\bigwedge^*(\mathbb{R}^m)^*$.
Here $\bigwedge^*(\mathbb{R}^m)^*$ is the vector space spanned by
$\bigl\{ v_1 \wedge \cdots \wedge v_l |v_{i=(1, \dots, l)} \in  \mathbb{R}^m ,~~ l\leqq m  \bigr\}$.

 For the operator representation, now we reexpress the commutation relation of the super Heisenberg algebra (\ref{cmsplie})
 with new generators $X_i, ~ Y_i, ~ \theta_a^1, ~ \theta_a^2 $ and $I$
 by the following map
 \begin{equation}
 \label{opmapsp}
 q_i \longrightarrow i X_i , ~ p_i  \longrightarrow i Y_i, ~
 \xi_a \longrightarrow i \theta_a^1 , ~ \lambda_a  \longrightarrow i \theta_a^2,
 ~ r \longrightarrow -i  I .
 \end{equation}
 Then we have
\begin{equation}
\label{cmheis}
[X_i, Y_j] = i \delta_{ij} I ,~~ \{\theta_a^1 , \theta_b^2\} = i \delta_{ab} I, ~~ {\rm others}=0.
\end{equation}
A representation for $X_i, ~ Y_i, ~ \theta_a^1, ~ \theta_a^2 , $ and $I$ for operators can be given by
\begin{eqnarray}
 & &(X_i f)(s, \eta)  =  s_i f(s,\eta),~~~(Y_i f)(s,\eta)  =  -i\frac{\partial f}{\partial s_i}(s,\eta),  \nonumber \\
  & &(\theta_a^1 f)(s,\eta)  =  \eta_a f(s,\eta), ~~~ (\theta_a^2 f)(s,\eta) = i \frac{\partial f}{\partial \eta_a}(s,\eta) , \\
  & & I = {\rm identity}, \nonumber
 \end{eqnarray}
where
$s=(s_1,\ldots,s_n), ~\eta = (\eta_1,\ldots,\eta_m)$, and $(s,\eta)$ belongs to $\mathbb{R}^{n|m}$.
Note that here again $X_i,~ Y_i,~ \theta_a^1,~ \theta_a^2$ can be regarded  as the coordinates
of a non(anti)commutative superspace.

 For the first operator representation, we send $(t,x,y,\alpha,\beta)$ to
$\chi(t)\varepsilon(x,\alpha)\delta(y,\beta)$,
where $(x,\alpha), (y,\beta) \in \mathbb{R}^{n|m}$:
\begin{eqnarray}
\chi(t) :  ~ f(s,\eta) & \longrightarrow & e^{- i t } f(s,\eta), \nonumber \\
\varepsilon(x,\alpha):  ~ f(s,\eta) & \longrightarrow &
e^{ i (x \cdot X + \alpha \cdot \theta^1) }f(s,\eta)
 = e^{ i (x \cdot s + \alpha \cdot \eta) }f(s,\eta) ,  \\
\delta(y,\beta): ~ f(s,\eta) & \longrightarrow & e^{i (y \cdot Y +
\beta \cdot \theta^2) } f(s,\eta) = f(s+y,\eta+\beta) . \nonumber
\end{eqnarray}

 For the second operator representation, we send $(t,x,y,\alpha,\beta)$ to
$U(t,x,y,\alpha,\beta)=
\chi(t+\frac{xy+\alpha\beta}{2})\varepsilon(x,\alpha)\delta(y,\beta)$.
As in the bosonic case, $U(t,x,y,\alpha,\beta)$ can be rewritten as
\begin{equation}
U(t,x,y,\alpha,\beta)=\exp\bigl[i(x\cdot X+y\cdot Y+\alpha \cdot \theta^1
+\beta \cdot \theta^2 -t I )\bigr].
\label{repr}
\end{equation}
 From now on,  for the sake of brevity we will drop $I$ for the identity,
and will use the above form (\ref{repr}) in the following section.

Supersymmetric extensions of the Stone-von Neumann theorem were
considered in \cite{gp88,lo94}.
It was shown in \cite{lo94} that there exists a unique irreducible
unitary $sH(2n|2m)$ module up to isomorphism described as  in the bosonic case.
\\

\section{Deformed superspace}\label{spspace}


\subsection{Deformed $\mathcal{N}=(2,2)$ superspace in two dimensions}

\noindent
 First we introduce the two-dimensional $\mathcal{N}=(2,2)$
superspace spanned by $(X^{\mu}, \theta^{\alpha},
\bar{\theta}^{\dot{\alpha}})$. $\theta$ and $\bar{\theta}$ are
transformed as a spinor and a conjugate spinor, respectively.
We deform this superspace by the following
commutation relations \cite{Se,BeSe}:
\begin{align}
[X^{1}, X^{2}]&=i\varTheta-2iC\bar{\theta}^{1}\bar{\theta}^{2},&
[X^{1}, \theta^{1}]&=iC\bar{\theta}^{2},&
[X^{1}, \theta^{2}]&=iC\bar{\theta}^{1},
\notag\\
[X^{2}, \theta^{1}]&=C\bar{\theta}^{2},&
[X^{2}, \theta^{2}]&=C\bar{\theta}^{1},&
\{\theta^{1}, \theta^{2}\}&=C,\label{2dcr1}
\end{align}
where $\bar{\theta}$'s commute or anticommute with other coordinates,
\textit{i.e.} the center. $\varTheta$ and $C$ are constants.
In the case of $\varTheta=0$, \eqref{2dcr1} is obtained by
dimensional reduction of the commutation relation of non(anti)commutative
$\mathcal{N}=1$ superspace in four dimensions.%
\footnote{
In four dimensions, there are three non(anti)commutative parameters
$C^{11}$, $C^{12}$ and $C^{22}$ in general.
The parameter $C$ in \eqref{2dcr1} corresponds to $C^{12}$. We have set
$C^{11}=C^{22}=0$ for simplicity.
}
On the other hand, in the case of $C=0$, \eqref{2dcr1} reproduces
the two-dimensional noncommutative space (with usual fermionic coordinates).
Although we do not derive \eqref{2dcr1} from superstring,
we use \eqref{2dcr1} as a simple unified expression of above two cases.
The Moyal product corresponding to \eqref{2dcr1} is given by
\begin{align}
&F(X,\theta,\bar{\theta})\ast G(X,\theta,\bar{\theta})=\notag\\
&\exp\biggl[\frac{i}{2}\varTheta(\partial_{1}\partial'_{2}
-\partial_{2}\partial'_{1})-\frac{1}{2}C(\mathcal{Q}_{1}\mathcal{Q}'_{2}
+\mathcal{Q}_{2}\mathcal{Q}'_{1})\biggr]
F(X,\theta,\bar{\theta})G(X^{\prime},\theta',\bar{\theta}')
\biggr|_{(X^{\prime\mu},\theta',\bar{\theta}')=(X^{\mu},\theta,\bar{\theta})},
\label{Moyal}
\end{align}
where $\partial_{\mu}=\partial/\partial X^{\mu}$.
$\mathcal{Q}_{1},\,\mathcal{Q}_{2}$ are the supercharges defined by
\begin{equation}
\mathcal{Q}_{1}=\frac{\partial}{\partial\theta^{1}}
-i\bar{\theta}^{1}(\partial_{1}+i\partial_{2}),\quad
\mathcal{Q}_{2}=\frac{\partial}{\partial\theta^{2}}
-i\bar{\theta}^{2}(\partial_{1}-i\partial_{2}).
\end{equation}
$\partial'_{\mu},\,\mathcal{Q}'_{1},\,\mathcal{Q}'_{2}$
are respectively obtained from
$\partial_{\mu},\,\mathcal{Q}_{1},\,\mathcal{Q}_{2}$ by replacement
$(X^{\mu},\theta,\bar{\theta})\to(X^{\prime\mu},\theta',\bar{\theta}')$.
In general, from the consistency with supersymmetry, superspace can be deformed
in terms of Moyal product which includes either supercharges $\mathcal{Q}_{\alpha}$
or supercovariant derivatives $D_{\alpha}$ in order to obtain non(anti)commutativity
in fermionic coordinates, where $D_{\alpha}$ is defined by
\begin{equation}
D_{1}=\frac{\partial}{\partial\theta^{1}}
+i\bar{\theta}^{1}(\partial_{1}+i\partial_{2}),\quad
D_{2}=\frac{\partial}{\partial\theta^{2}}
+i\bar{\theta}^{2}(\partial_{1}-i\partial_{2}).
\end{equation}
Now $\mathcal{Q}_{\alpha}$ is used in the Moyal product as in \eqref{Moyal}.
In this case,
the half of the supersymmetry is broken but the chirality of superfields is preserved.
If $D_{\alpha}$ is used in the Moyal product instead of $\mathcal{Q}_{\alpha}$,
the supersymmetry is fully preserved but the chirality of superfields
is broken \cite{Klemm:2001yu,Se}.

Now we change the normalization of the operators by
\begin{equation}
X^{\mu}\to \sqrt{\varTheta}\,X^{\mu},\quad
\theta^{\alpha}\to \sqrt{iC}\,\theta^{\alpha},\quad
\bar{\theta}^{\dot{\alpha}}\to \sqrt{\frac{\varTheta}{iC}}\,
\bar{\theta}^{\dot{\alpha}}.
\end{equation}
Then the commutation relations \eqref{2dcr1} become
\begin{align}
[X^{1}, X^{2}]&=i-2\bar{\theta}^{1}\bar{\theta}^{2},&
[X^{1}, \theta^{1}]&=\bar{\theta}^{2},&
[X^{1}, \theta^{2}]&=\bar{\theta}^{1},
\notag\\
[X^{2}, \theta^{1}]&=-i\bar{\theta}^{2},&
[X^{2}, \theta^{2}]&=-i\bar{\theta}^{1},&
\{\theta^{1}, \theta^{2}\}&=-i,\label{2dcr2}
\end{align}
As in the bosonic case, we can introduce
a unique operator representation of the above algebra
as in  (\ref{repr}),
as a representative of the corresponding
super Heisenberg group
\begin{equation}
U_{A}=
\exp\Bigl[i\bigl(x_{A}X^{1}+y_{A}X^{2}+\alpha_{A}\theta^{1}+\beta_{A}\theta^{2}
-\tilde{t}_{A}
\bigr)\Bigr].
\label{U}
\end{equation}
This satisfies
\begin{align}
\label{Ucom}
U_{A}U_{B}&=\exp(-i\varXi_{AB})U_{B}U_{A},\\[2mm]
\varXi_{AB}&=
(1+2i\bar{\theta}^{1}\bar{\theta}^{2})(x_{A}y_{B}-x_{B}y_{A})
+(\alpha_{A}\beta_{B}-\alpha_{B}\beta_{A})\notag\\
&\qquad\quad{}+i(x_{A}-iy_{A})\beta_{B}\bar{\theta}^{1}
-i(x_{B}-iy_{B})\beta_{A}\bar{\theta}^{1}\notag\\
&\qquad\quad{}-i(x_{A}+iy_{A})\alpha_{B}\bar{\theta}^{2}
+i(x_{B}+iy_{B})\alpha_{A}\bar{\theta}^{2}\Bigr].
\end{align}
%

Now, we can give two types of corresponding supermatrix representations.
One is the $4 \times 4$ supermatrix representation as
\begin{align}
&\pi_{A}=
\exp\bigl[M(\tilde{t}_{A},x_{A},y_{A},\alpha_{A},\beta_{A})\bigr],\\[2mm]
&M(\tilde{t}_{A},x_{A},y_{A};\alpha_{A},\beta_{A})=
\left(
\begin{array}{ccc|c}
0 & x_{A} & \tilde{t}_{A} &\tilde{\alpha}_{A} \\
0 & 0 & y_{A} & 0 \\
0 & 0 & 0 & 0 \\
\hline
0 & 0 & \tilde{\beta}_{A} & 0
\end{array}
\right),\label{44}
\end{align}
where
\begin{equation}
\tilde{\alpha}_{A}=\alpha_{A}-i(x_{A}-iy_{A})
\bar{\theta}^{1},\quad
\tilde{\beta}_{A}=\beta_{A}+i(x_{A}+iy_{A})
\bar{\theta}^{2}.
\end{equation}
In order to obtain \eqref{44}, It is convenient
to use the chiral coordinate $Y^{\mu}$
which is defined by
\begin{equation}
Y^{1}=X^{1}+i\theta^{1}\bar{\theta}^{1}-i\theta^{2}\bar{\theta}^{2}, \quad
Y^{2}=X^{2}+\theta^{1}\bar{\theta}^{1}+\theta^{2}\bar{\theta}^{2}.
\end{equation}
The commutation relation of $Y^{\mu}$ and $\theta^{\alpha}$ is
\begin{equation}
[Y^{1},Y^{2}]=i, \quad [Y^{\mu},\theta^{\alpha}]=0,\quad
\{\theta^{1}, \theta^{2}\}=-i. \label{y}
\end{equation}
In terms of $Y^{\mu}$, \eqref{U} is rewritten as
\begin{equation}
U_{A}=
\exp\Bigl[i\bigl(x_{A}Y^{1}+y_{A}Y^{2}
+\tilde{\alpha}_{A}\theta^{1}+\tilde{\beta}_{A}\theta^{2}
-\tilde{t}_{A}\bigr)\Bigr].
\label{U2}
\end{equation}
 From \eqref{y} and \eqref{U2},
we obtain the representation \eqref{44}.
In this representation,
we regard $\bar{\theta}^{1}$ and $\bar{\theta}^{2}$
as grassmann numbers.

 For the other type of matrix representation,
 we regard
$\bar{\theta}^{1}$ and $\bar{\theta}^{2}$
as the operators which belong to
the center of the super Heisenberg group.
The central element $\tilde{t}_{A}$ in \eqref{U} is replaced with
\begin{equation}
\tilde{t}_{A}
\rightarrow
t_{A}+\bar{\alpha}_{A}\bar{\theta}^{1}+\bar{\beta}_{A}\bar{\theta}^{2}
+u_{A}\bar{\theta}^{1}\bar{\theta}^{2},
\label{U3}
\end{equation}
where $t_{A},\,\bar{\alpha}_{A},\,\bar{\beta}_{A},\,u_{A}$ are the parameters
corresponding to the center of the super Heisenberg group
$I,\,\bar{\theta}^{1},\,\bar{\theta}^{2},\,\bar{\theta}^{1}\bar{\theta}^{2},$
respectively.
In this case, the representation is given by
the $6 \times 6$ supermatrix $\tilde{\pi}_{A}$ as
\begin{align}
&\tilde{\pi}_{A}=\exp\bigl[\tilde{M}(t_{A},x_{A},y_{A},u_{A},
\alpha_{A}, \beta_{A}, \bar{\alpha}_{A}, \bar{\beta}_{A})
\bigr],\\[2mm]
&\tilde{M}(t_{A},x_{A},y_{A},u_{A},
 \alpha_{A}, \beta_{A}, \bar{\alpha}_{A}, \bar{\beta}_{A})
=\notag\\
&\qquad\qquad\qquad\qquad\left(
\begin{array}{ccc|ccc}
0 & x_{A} & t_{A} &\alpha_{A} & 0 & \bar{\beta}_{A} \\
0 & 0 & y_{A} & 0 & 0 & 0 \\
0 & 0 & 0 & 0 & 0 & 0 \\
\hline
0 & 0 & \beta_{A} & 0 & 0 & -i(x_{A}+iy_{A}) \\
0 & 0 & \bar{\alpha}_{A} & i(x_{A}-iy_{A})& 0 &
u_{A} \\
0 & 0 & 0 & 0 & 0 & 0
\end{array}
\right).
\end{align}
In terms of $\tilde{\pi}_{A}$, as in \eqref{multi},
we have the multiplication rule
\begin{multline}
(t_{A}, x_{A}, y_{A}, u_{A},
\alpha_{A}, \beta_{A}, \bar{\alpha}_{A}, \bar{\beta}_{A})
\cdot
(t_{B}, x_{B}, y_{B}, u_{B},
\alpha_{B}, \beta_{B}, \bar{\alpha}_{B}, \bar{\beta}_{B})
=\\
(t_{AB}, x_{A}+x_{B}, y_{A}+y_{B}, u_{AB},
\alpha_{A}+\alpha_{B}, \beta_{A}+\beta_{B},
\bar{\alpha}_{AB}, \bar{\beta}_{AB}).
\end{multline}
Here $t_{AB}$, $u_{AB}$, $\bar{\alpha}_{AB}$ and $\bar{\beta}_{AB}$
are given by
\begin{align}
t_{AB}&=t_{A}+t_{B}+\frac{1}{2}(x_{A}y_{B}-x_{B}y_{A})
+\frac{1}{2}(\alpha_{A}\beta_{B}-\alpha_{B}\beta_{A}), \nonumber
\\
u_{AB}&=u_{A}+u_{B}+i(x_{A}y_{B}-x_{B}y_{A}), \nonumber
\\
\bar{\alpha}_{AB}&=\bar{\alpha}_{A}+\bar{\alpha}_{B}
+\frac{i}{2}(x_{A}-iy_{A})\beta_{B}-\frac{i}{2}(x_{B}-iy_{B})\beta_{A}, \nonumber
\\
\bar{\beta}_{AB}&=\bar{\beta}_{A}+\bar{\beta}_{B}
-\frac{i}{2}(x_{A}+iy_{A})\alpha_{B}+\frac{i}{2}(x_{B}+iy_{B})\alpha_{A}.
\end{align}
The commutation relation among $\tilde{\pi}$'s is the same as that of
$U$'s in (\ref{Ucom}):
\begin{align}
\tilde{\pi}_{A}\tilde{\pi}_{B}&=\exp(\Omega_{AB})\tilde{\pi}_{B}\tilde{\pi}_{A},
\end{align}
where
\begin{align}
&\Omega_{AB}
 = \left(
\begin{array}{ccc|ccc}
0 & 0 & t_{[AB]} & 0 & 0 & \bar{\beta}_{[AB]} \\
0 & 0 & 0 & 0 & 0 & 0 \\
0 & 0 & 0 & 0 & 0 & 0 \\
\hline
0 & 0 & 0 & 0 & 0 & 0 \\
0 & 0 & \bar{\alpha}_{[AB]} & 0& 0 &
u_{[AB]} \\
0 & 0 & 0 & 0 & 0 & 0
\end{array}
\right)
\end{align}
and the bracket means antisymmetrization of indices.


\subsection{Toy superspace with non-central extension}

Super Heisenberg group is the central extension of the ordinary
superspace. However, some of noncommutative superspaces
which cannot be obtained by central extension admit
the matrix representation similar to super Heisenberg group.
In this section, we consider such an example
of the superspace with non-central extension.

Here we start from the module which is given by $f(s,\zeta,\bar{\zeta})$,
where $(s,\zeta,\bar{\zeta})$ is the coordinates of ordinary one-dimensional
superspace. From this module, we define the coordinates of the
two-dimensional deformed superspace by%
\footnote{Here we have already changed the normalization of the operators
to absorb noncommutative and non-anticommutative parameters
as in the case of \eqref{2dcr2}.}
\begin{align}
\hat{X}^{1}=s\cdot{},\quad \hat{X}^{2}=-i\frac{\partial}{\partial s},\quad
\hat{\theta}^{1}=-i\zeta\cdot{},\quad
\hat{\bar{\theta}}^{1}=-i\bar{\zeta}\cdot{},\quad
\hat{\theta}^{2}=
\frac{\partial}{\partial \zeta}+\bar{\zeta}\frac{\partial}{\partial s},\quad
\quad \hat{\bar{\theta}}^{2}=
\frac{\partial}{\partial \bar{\zeta}}+\zeta\frac{\partial}{\partial s}.
\end{align}
The nontrivial commutation relations are
\begin{align}
\bigl[\hat{X}^{1},\hat{X}^{2}\bigr]&=i,&
\bigl[\hat{X}^{1},\hat{\theta}^{2}\bigr]&=-i\hat{\bar{\theta}}^{1},&
\bigl[\hat{X}^{1},\hat{\bar{\theta}}^{2}\bigr]&=-i\hat{\theta}^{1},
\notag\\
\bigl\{\hat{\theta}^{1},\hat{\theta}^{2}\}&=-i,&
\bigl\{\hat{\bar{\theta}}^{1},\hat{\bar{\theta}}^{2}\}&=-i,&
\bigl\{\hat{\theta}^{2},\hat{\bar{\theta}}^{2}\}&=2i\hat{X}^{2}.&
\label{nce}
\end{align}
The last eqation in \eqref{nce} resembles the deformed superspace defined in
\cite{de Boer:2003dn}.
Now the commutators and
the anticommutators contain the operators which are not centers. Then this superspace
is not a central extension of the ordinary two-dimensional superspace\footnote{
However the bosonic part of this algebra is Heisenberg algebra.
}.
Despite this, the matrix representation of \eqref{nce} can be constructed
as in the case of super Heisenberg group.
%
%
%
We define $\hat{U}_{A}$ as follows like the representative of
super Heisenberg group in the previous subsection
\begin{equation}
\hat{U}_{A}=
\exp\bigl[i(\hat{x}_A\hat{X}^{1}+\hat{y}_A\hat{X}^{2}
+\hat{\alpha}_A\hat{\theta}^{1}+\hat{\beta}_A\hat{\theta}^{2}
+\hat{\bar{\alpha}}_A\hat{\bar{\theta}}^{1}
+\hat{\bar{\beta}}_A\hat{\bar{\theta}}^{2}-\hat{t}_{A})\bigr].
\label{Unc}
\end{equation}
This satisfies
\begin{align}
\label{cruh}
\hat{U}_{A}\hat{U}_{B}&=\exp(-i\hat{\varXi}_{AB})\hat{U}_{B}\hat{U}_{A},
\end{align}
where $\hat{\varXi}_{AB}$ is given by
\begin{align}
\hat{\varXi}_{AB}&=(\hat{x}_{A}\hat{y}_{B}-\hat{y}_{A}\hat{x}_{B})
+(\hat{\alpha}_{A}\hat{\beta}_{B}+\hat{\beta}_{A}\hat{\alpha}_{B})
+(\hat{\bar{\alpha}}_{A}\hat{\bar{\beta}}_{B}
+\hat{\bar{\beta}}_{A}\hat{\bar{\alpha}}_{B})
\notag\\
&\qquad{}
+\frac{3}{2}(\hat{\beta}_{A}\hat{\bar{\beta}}_{B}
+\hat{\bar{\beta}}_{A}\hat{\beta}_{B})(\hat{x}_{A}+\hat{x}_{B})
-(\hat{x}_{A}\hat{\bar{\beta}}_{B}
-\hat{\bar{\beta}}_{A}\hat{x}_{B})\hat{\theta}^{1}
\notag\\
&\qquad{}
-(\hat{x}_{A}\hat{\beta}_{B}-\hat{\beta}_{A}\hat{x}_{B})\hat{\bar{\theta}}^{1}
-2(\hat{\beta}_{A}\hat{\bar{\beta}}_{B}
+\hat{\bar{\beta}}_{A}\hat{\beta}_{B})\hat{X}^{2}.
\end{align}
%
We can assign the corresponding matrix representation of $\hat{U}_A$ as the supermatrix $\hat{\pi}_A$
\begin{align}
&\hat{\pi}_A=\exp\bigl[\hat{M}(\hat{t}_A, \hat{x}_A,\hat{y}_A,
\hat{\alpha}_A, \hat{\bar{\alpha}}_A,\hat{\beta}_A, \hat{\bar{\beta}}_A)\bigr],
\end{align}
where $\hat{M}$ is a $5 \times 5$ supermatrix given by
\begin{align}
&\hat{M}(\hat{t}_A, \hat{x}_A,\hat{y}_A,
\hat{\alpha}_A, \hat{\bar{\alpha}}_A,\hat{\beta}_A, \hat{\bar{\beta}}_A)=
\left(
 \begin{array}{ccc|cc}
 0 & \hat{x}_A & \hat{t}_A &\hat{\alpha}_A & \hat{\bar{\alpha}}_A \\
 0 & 0 & \hat{y}_A & \hat{\bar{\beta}}_A & \hat{\beta}_A \\
 0 & 0 & 0 &0&0 \\ \hline
 0 & 0 & \hat{\beta}_A & 0 & 0 \\
 0 & 0 & \hat{\bar{\beta}}_A & 0 & 0
 \end{array}
\right).
\end{align}
It is straightforward to check that $\hat{\pi}$ satisfies the same
commutation relation (\ref{cruh}) of $\hat{U}$:
\begin{align}
\hat{\pi}_{A}\hat{\pi}_{B}&=\exp(\hat{\Omega}_{AB})\hat{\pi}_{B}\hat{\pi}_{A},
\end{align}
where
\begin{equation}
\hat{\Omega}_{AB} =
\left(
 \begin{array}{ccc|cc}
 0 & 0 & \hat{t}_{[AB]} & \hat{\alpha}_{[AB]} & \hat{\bar{\alpha}}_{[AB]} \\
 0 & 0 & \hat{y}_{[AB]} & 0 & 0 \\
 0 & 0 & 0 &0&0 \\ \hline
 0 & 0 & 0 & 0 & 0 \\
 0 & 0 & 0 & 0 & 0
 \end{array}
\right)
\end{equation}
with $ \hat{t}_{[AB]},~\hat{\alpha}_{[AB]},~\hat{\bar{\alpha}}_{[AB]},~\hat{y}_{[AB]}$
given as follows:
\begin{align}
\hat{t}_{[AB]}&=(\hat{x}_{A}\hat{y}_{B}-\hat{y}_{A}\hat{x}_{B})
+(\hat{\alpha}_{A}\hat{\beta}_{B}+\hat{\beta}_{A}\hat{\alpha}_{B})
+(\hat{\bar{\alpha}}_{A}\hat{\bar{\beta}}_{B}
+\hat{\bar{\beta}}_{A}\hat{\bar{\alpha}}_{B})
\notag\\
&\quad{}
+\frac{3}{2}(\hat{\beta}_{A}\hat{\bar{\beta}}_{B}
+\hat{\bar{\beta}}_{A}\hat{\beta}_{B})(\hat{x}_{A}+\hat{x}_{B}), \nonumber
\\
\hat{\alpha}_{[AB]}&=\hat{x}_{A}\hat{\bar{\beta}}_{B}
-\hat{\bar{\beta}}_{A}\hat{x}_{B}, \nonumber
\\
\hat{\bar{\alpha}}_{[AB]}&=
\hat{x}_{A}\hat{\beta}_{B}-\hat{\beta}_{A}\hat{x}_{B}, \nonumber
\\
\hat{y}_{[AB]}&=2(\hat{\beta}_{A}\hat{\bar{\beta}}_{B}
+\hat{\bar{\beta}}_{A}\hat{\beta}_{B}).
\end{align}
This confirms that we can use the above matrix representation
instead of dealing with more complicated operator
manipulations.
\\


%
\section{Conclusion}

In this paper, we construct the super Heisenberg group and
corresponding supermatrix representation by extending the result
known in the bosonic case.

The low-energy effective theory of D-branes in a background NSNS
$B$-field becomes the noncommutative field theory, and the algebra
of the coordinates becomes Heisenberg algebra.
 From Heisenberg algebra, Heisenberg group can be constructed by exponential mapping.
The matrix representation of Heisenberg group is useful to
construct noncommutative tori and quantum theta-functions.
One can construct noncommutative tori in a easier manner via the
embedding of the corresponding Heisenberg groups.

When the background RR field is turned on, the low-energy
effective theory of D-branes becomes the field theory on
non(anti)commutative superspace of which the fermionic coordinates
have nontrivial commutation relations, and super Heisenberg group
would appear.
 For the anologue of bosonic Heisenberg group, the
supermatrix representation of super Heisenberg group can be
constructed.
We explicitly carry out this construction by extending the
relation known in the bosonic case to the super case:
 the two-dimensional deformed $\mathcal{N}=(2,2)$
superspace containing non(anti)commutativity in both
bosonic and fermionic coordinates.
As in the bosonic case, the supermatrix representation of super
Heisenberg group would be useful to construct noncommutative
supertori and quantum super theta-functions \cite{EeKiNa}.

Furthermore, this supermatrix representation is also applicable to
deformed superspaces corresponding to non-centrally extended `super
Heisenberg groups'.
We demonstrate this construction with a toy model of two-dimensional
deformed superspace at the end.
\\


\vspace{5mm}

\noindent
{\Large \bf Acknowledgments}

\vspace{5mm} \noindent The authors thank KIAS for hospitality
during the time that this work was done. This work was supported
by Korean Council for University Education, grant funded by Korean
Government(MOEHRD) for 2006 Domestic Faculty Exchange (E. C.-Y.),
and by KOSEF Research Grant No. R01-2006-000-10638-0 (H. K.).
H.~N. is supported by
the Postdoctoral Research Program of Sungkyunkwan University (2007)
and this work is the result of research activities (Astrophysical Research
Center for the Structure and Evolution of the Cosmos (ARCSEC))
supported by KOSEF.
\\




\end{document}